\documentstyle[aps,epsf]{revtex}
    \def \be   {\begin{equation}}
\def \ee   {\end{equation}}
\def \l {\label}
\begin{document}
\input epsf
\title{Gauss vs Coulomb and the cosmological mass-deficit problem}
\author{Manoelito M de Souza and Robson N. Silveira}
\address{Universidade Federal do Esp\'{\i}rito Santo - Departamento de
F\'{\i}sica\\29065.900 -Vit\'oria-ES-Brasil}
\date{\today}
\maketitle
\begin{abstract}
In a previous work General Relativity has been presented as a microscopic theory of finite and discrete point-like fields that we associate to a classical description of gravitons. The standard macroscopic continuous field is retrieved as an average-valued field through an integration over these gravitons. Here we discuss extreme alternative (the Gauss's and the Coulomb's) ways of obtaining and interpreting the averaged fields, how they depend on the kind of measurements involved, and how do they fit with the experimental data. The field measurements in the classical tests of general relativity correspond to the Coulomb's mode whereas the determination of the overall spacetime curvature in a cosmological scale is clearly a Gauss's mode. As a natural consequence there is no missing mass and, therefore, no such a need of dark mass as the value predicted by General Relativity, in the context of the Gauss's mode, agrees with the observed one.
\end{abstract}
\begin{center}
\end{center}

\section{Introduction}
\noindent This note is a sequence to the references \cite{hep-th/9610028,gr-qc/9801040} where photons and gravitons are described as discrete and finite classical fields.  Our objective here is to present and discuss the connections between field measurements and their alternative interpretations with regard to the concept of discrete gravitational field developed in \cite{gr-qc/9801040}. It is  related to the question whether  the field of a single point source exist independently  of the presence of a test-body.
How far does a classical field actually represents the experimentally observed interactions? 
This is closely connected to the distinct contents of the Coulomb's law and of the Gauss's law. While the first one gives a strict description of what is actually observed, i.e., a force between two charges, acting on each one along the straight line connecting them, the second one contains an extra assumption (the Faraday-Maxwell concept of field) that effectively {\bf extends} this effect, observed at the charge position only, to all points in the space around the charge, regardless the presence or not of the probe charge. It is worth repeating: the concept of a field existing everywhere around a single charge, regardless the presence of any other charge, is an extrapolation of what is effectively observed.\\
There is, therefore, a very deep distinction between the Coulomb's and the Gauss's laws. The last one describes the {\it inferred} electric field as existing around a single charge, independently of the presence of the other charge.\\
This question has been taken as irrelevant to physics. It belongs to the realm of metaphysics as it has been considered that experimentally it is impossible to distinguish between the two views (Gauss's and Coulomb's) as the field is only indirectly determined and it necessarily requires the use of a test-body. But in the context of a discrete field formalism it acquires high and new relevance.  In the standard continuous formalism a field is a function defined with a lightcone support and, naturally, represents what we are calling the Gauss's view. In the discrete field formalism the Coulomb's view is the most appropriate one as the field is a spacetime-point disturbance defined and propagating without a change on a straight line, on a generator of the lightcone which means that its momentum is also fixed. It is a point in the phase space and, as such, it has a finite number of degrees of freedom and that is why it generates a finite formalism.  As a solution to the field equations, the physical field exists only on the straight-line segment connecting two interacting point-sources and there is field only in the presence of the test body. A single isolated source has no field. The discrete field, in contradistinction to the standard continuous one has no singularity. 

The Coulomb's mode of interpreting the gravitational field as existing only on the line connecting the two interacting point masses reminds the hypothesized flux tube of gluons of Quantum Chromodynamics between an interacting  quark-antiquark pair, which would be responsible for the quark confinement. This connection sounds quite interesting considering the difficulties on trying to have quark-confinement proved in QCD.\\

The standard continuous field, as discussed in \cite{gr-qc/9801040} is retrieved by a replacement of this, let's say, real quantum (in the sense of discrete smallest part of a) field, by an integration over a continuous isotropic distribution of longitudinal fields. It necessarily contains these non-physical components. They are the ones that, must be excluded by, for example, some gauge restriction on the state space, in a field quantization procedure.\\
Why is the Gauss's vs Coulomb's dilemma now relevant and how can it be experimentally settled down? For the electromagnetic field there is, effectively, no way of resolving it, and so, there it remains a physically irrelevant question. A very distinct situation occurs with the gravitational field, since its tensorial character and its association to the metric tensor in general relativity allows that the validity of the inference implicit in the Gauss's law be experimentally checked. A point-field $A_{f}(x)$ defined on a straight line can be seen, for simplicity, as the standard continuous field $A(x)$, defined on a lightcone, restricted to its intersection with the lightcone generator tangent to the light-like four-vector $f$, which is defined by the two simultaneous restrictions on a $\Delta x$ associated to the field propagation:
\be
\cases{\Delta\tau^2+\Delta x^{2}=0& (a lightcone, for $\Delta\tau=0$ )\cr 
	\Delta\tau+f.\Delta x=0&(the hyperplane tangent to the lightcone),\cr}
\ee
or just by $\Delta x.(\eta+ff).\Delta x=0$, where $\eta$ is the Minkowski metric $diag.(1,1,1,-1).$ We are adopting here the same notations and conventions of \cite{gr-qc/9801040}.
In general relativity with extended causality the solution of the Einsteins's field equations in vacuum, $R^f_{\mu\nu}=0$, for the metric tensor field $g^{f}_{\mu\nu}$ restricted to a straight line tangent to $f$,

\be
g^{f}_{\mu\nu}=g_{\mu\nu}{\big |}_{\Delta x.(\eta+ff).\Delta x=0}
\ee

is given by
\be
\l{gtf}
g^{f}_{\alpha\beta}(t,r,\theta,\varphi)=\cases{\eta_{\alpha\beta},& for $\theta\ne\theta_{f},\; \varphi\ne\varphi_{f}$;\cr
\eta_{\alpha\beta}+\chi\delta(\tau+f.x)f_{\alpha}f_{\beta}& for $\theta=\theta_{f},\; \varphi=\varphi_{f}$,\cr}
\ee
or explicitly 
\begin{equation}
\label{recSch}
\text{ }g^{f}_{\alpha \beta }(t,r,\theta_{f} ,\varphi_{f} )=\pmatrix{
-1 & 0 & 0 & 0 \cr 
 0 & 1 & 0 & 0 \cr 
 0 & 0 & r^2 & 0 \cr 
 0 & 0 & 0 & r^2\sin^2\theta_{f}\cr} +\chi\delta(f.\Delta x)\pmatrix{
{ f}_0^2 & { f}_0{ f}_r & 0 & 0 \cr 
{ f}_r{ f}_0 & { f}_r^2 & 0 & 0 \cr 
0 & 0 & 0 & 0 \cr 
0 & 0 & 0 & 0\cr},
\end{equation}
with  $f_{\mu}=(f_0,\ f_r,0,0),$ and with $\theta_{f}$ and $\varphi _{f} ,$ defining the space direction  $\vec{f}$ of $f$. The metric $\,\,\ g^{f}_{\alpha \beta }$
represents a single, let's say, ``classical quantum" of gravity  propagating along a line $f$  and
observed as an event $(t,r,\theta _{f},\varphi _{f})$ at the probe mass. Let us, in an abuse of language,  call it the graviton on the fibre $f,$ for shortness. 
As we learned from \cite{gr-qc/9801040}, for retrieving the standard Schwarzschild solution we must replace the single real graviton propagating on the ray $(\theta_{f},\varphi _{f})$ by a continuous isotropic distribution of virtual gravitons $g_{f'}$ and then integrate over all directions $(\theta_{f'},\varphi _{f'})$. We write $f'.x=t-r\cos\theta_{f'}$ by putting our z-coordinate axis on the line passing by the probe-mass. We take $f'^4=|{\vec f'}|=1,$
expressing \cite{gr-qc/9801040} that we are in the  instantaneous rest frame of the field source at the time of emission of $g_{f}$.
\begin{equation}
\int d\varphi _f'\ sen\theta _f'\ d\theta _f'\ \delta (t-|\vec r|\cos \theta _f')=\frac{2\pi }r\int_{-1}^{1}d\cos
\theta _f'\,\delta (\cos \theta _f'-\frac{t}r)=
\cases{\frac{2\pi }r,&for $t\in[0,r]$;\cr
0,&for $t\notin[0,r]$.\cr}
\end{equation}
This condition on $t$ means that the deformation on the flat spacetime that we are associating to a graviton is not null as far as $t$ is smaller, or at least, equal to $\frac rc$, the time that the graviton, after being emitted by the source at the origin, takes to reach the probe mass at $(t,r,\theta_{f},\phi_{f})$, where it is absorbed. This process is continued for $t>\frac rc$  by other gravitons subsequently emitted. So, the large number of gravitons emitted (and absorbed) in any realistic experiment transmit the idea of continuity and of a static field. Thus we can write for $\theta=\theta _f$ and $\varphi=\varphi _f$ that 
\be
\l{gp}
ds(t,r,\theta _{f}\,\,,\varphi _{f})^2=-(1-\frac \chi
r)\,dt^2+(1+\frac \chi r)\,dr^2-\frac{2\chi}{r}dtdr+r^2d^2\Omega,
\ee
where $d^2\Omega=d\theta^2+sen^2\theta
\,d\varphi^2$.
\\A well-known \cite{Weinberg} simple coordinate transformation puts (\ref{gp}) in the standard  diagonal form
\begin{equation}
\l{ss}
ds^2=-(1-\frac \chi r)\,dt^2+(1-\frac \chi r)^{-1}\,dr^2+r^2d^2\Omega\qquad{\hbox{for }}\quad\theta=\theta _{f},\;\varphi=\varphi_{f},  \label{campoSch}
\end{equation}
This is the Schwarzschild solution with all its properties and physical meaning and consequences except with the distinguishing interpretation that the event $(t,r,\theta_{f},\phi_{f})$ refers to the probe-mass location. 
A probe mass m, wherever be it placed, detects a Schwarzschild metric on the spacetime around the mass $M=\frac{\chi}{2G}$ on the coordinate origin. The observer concludes then that M is the symmetry center of a Schwarzschild spacetime, but actually the spacetime is flat except  on the  straight-line segment connecting m to $M$; it is completely flat, except at the origin ($M$), in the absence of m. There is no problem of continuity because (\ref{ss})  is a continuous function of the probe-mass location and any differentiation of $g_{\mu\nu}$ corresponds to a small displacement of the probe mass. So, the Coulomb's view is the natural interpretation of the Schwarzschild solution.\\ 
Physics is theorization on measurement results. It is of a particular importance the association we make between the numbers we get in our measurements and the intellectual constructions we create to rationalize them.  In the context of a discrete field formalism there is no continuous interaction, It is just an approximation justified by the inertia of our observation means and by the enormous, with respect to our anthropic scales, density of interaction events. It means that the idea of acceleration, force and interaction fields as continuous functions as well as the use of calculus (integral and derivatives) are just useful approximations that we must know their limits of validity. Any idea of continuity, except that of the spacetime itself, is just an approximation which is valid only in very specific limits. The concept of elementary fields, of their sources and of their interactions as something continuously spread over some regions of spacetime  is included in such a category of useful but limited approximations. They are all conceived as discrete and point-like. Being elementary means being structureless, point-like. Observe that having spin, which we are not considering in this note, does not mean that the elementary object has structure but that it has components, i.e. it is described by a set of elementary point-like fields.\\
The probe-mass idea allows the connection between our theoretical idealizations and the results of our measurements. We consider as a good approximation if the probe-mass is small in comparison to a  mass scale characterizing the system under consideration (for not changing the field configuration), and if its size is small in comparison to its distance to the interaction sources. It can indeed be a point-object in the case of an elementary particle, like an electron for example, or it can be relatively large like a rock or a planet. So, although the Coulomb's mode corresponds strictly to a description of interactions between point objects it may be applied as a good approximation in certain cases for interactions between extended but treated as point objects.\\
We can conceive the existence of point-like elementary objects but we know that the experimental determination  of their position in space and in time cannot be better than the one given by the accuracy of our measuring apparatus. We are not talking about the Uncertainty Principle; we refer to the apparatus aperture, as a zero aperture would mean no measuring at all. Thus, measurements do not give us access to the actual properties of the discrete fields but just to a kind of their space-and-time averages. 

We can also force the generation of the Gauss's view. 
The Gauss's inference is that all directions are equivalent; the probe mass direction $(\theta_{f},\varphi _{f})$ has nothing of special as far as the field generated by the source at the origin is concerned. In this case, the integration $\int d^2\Omega_{f'}$ on the $f'$-space is replaced by the integration $\int d^2\Omega$ on  configuration space with ${\vec f'}=\frac{{\vec x}}{|\vec x|}=\frac{{\vec x}}{r}.$ We take for $t>0$, $$g(x,\tau)=\frac{1}{4\pi}\int df'^4 d^{2}\Omega_{f'}g_{f'}(x,\tau)$$ and $$\delta(f'.x)=\frac{1}{|t|}\delta(f'^4+\frac{{\vec f'}.{\vec x}}{t})=\frac{1}{|t|}\delta(f'^4+\frac{|{\vec x}|}{t}),$$ with $r^{2}=x^{2}+y^{2}+z^{2}=t^{2}.$ Then, we have
\be
g_{44}=\eta_{44}+\frac{\chi}{4\pi r}\int d^{2}\Omega_{f'}(f'^4)^2=\eta_{44}+\frac{\chi}{4\pi r}\int d^{2}\Omega \frac{r^2}{t^2}=-1+\frac{\chi}{r},
\ee
\be
g_{4i}=\frac{\chi}{4\pi r}\int d^{2}\Omega_{f'}f'^4f^{i}=-\frac{\chi}{4\pi r}\int d^{2}\Omega \frac{x^{i}r}{t^2}=0,
\ee
and 
\be
g_{ij}=\eta_{ij}+\frac{\chi}{4\pi r}\int d^{2}\Omega_{f'}f'^if'^j=\delta_{ij}+\frac{\chi}{4\pi r}\int d^{2}\Omega \frac{x^ix^j}{t^2}=(1-\frac{\chi}{3r})\delta_{ij}.
\ee
So, we can write it in the standard form
\be
\l{sls}
ds^2=-(1-\frac{\chi}{r})dt^2+(1-\frac{\chi}{3r})(dx^{2}+dy^{2}+dz^{2}),
\ee
which is a solution of the linearized Einstein's equations in vacuum. It is distinguishable from (\ref{gp}) and is similar to the linearized Schwarzschild metric in isotropic coordinates, from which it only differs for the $\slantfrac{1}{3}$ factor in the metric first-order space components. It is interesting that the Coulomb's view reproduces the generic solution of the full-fledged Einstein's vacuum equations whereas the Gauss's view reproduces just a solution of the linearized equations. Besides, (\ref{sls}) is not the linearized version of (\ref{ss}). So,  in contradistinction to the case of the Maxwell's theory in general relativity the Gauss's and the Coulomb's views correspond to two physically distinguishable macroscopic descriptions. (\ref{ss}) and (\ref{sls}) do represent two distinct solutions for a problem which is proved to have no more than one solution \cite{Birkhoff}. It means that they cannot be both applicable to a same physical situation. Their applicability depends on the kind of measurement (averaging) involved.\\

Let us use the pos-Newtonian analysis for comparing the performances of the two metrics (\ref{ss}) and (\ref{sls}) on fitting the classical experimental tests. In common, they can all be treated as point-mass--point-mass (two-body) interactions.
We can divide these tests in three kinds:
\begin{enumerate}
\item The tests that depend only on $g_{44}$. They do not distinguish between (\ref{ss}) and (\ref{sls}). The red shift of the stellar spectrum, for example.
\item Tests where the probe-body describes an integer number of turns around the source. The difference between  the effects of (\ref{ss}) and of (\ref{sls}) tends to be minimized making it harder to discriminate these metrics. The advance of planetary perihelia is an example.
\item Tests where the probe-body does not describe a complete turn around the source. In this case the average field must produce just a fraction (roughly $\frac{\Delta\theta}{2\pi}$) of the actual result.
\end{enumerate}
Considering the remarkably good\cite{Will} agreement of the predictions of general relativity with the experimental data we can just compare the Gauss's mode predictions with the respective ones from general relativity, which are coincident with the ones of the Coulomb's mode.
In terms of pos-Newtonian parameters, which are defined by 
\be
ds^2=-(1-2\alpha\frac{MG}{r}+2\beta\frac{M^2G^2}{r^2}+\cdots)dt^2+(1-2\gamma\frac{MG}{r}+\cdots)d\sigma^2,
\ee
 the advance of planetary perihelia is given by
\be
\Delta\phi=\frac{2-\beta+2\gamma}{3}\Delta\phi_{E},
\ee
where $\Delta\phi_{E}$ is the value predicted by general relativity, that is, with $\alpha=\beta=\gamma=1$. For the metric (\ref{sls}) we know that $\alpha=1$, $\beta=0$ and $\gamma=1/3.$  This produces
\be
\l{15}
\Delta\phi=\frac{8}{9}\Delta\phi_{E}.
\ee
This is the Gauss's mode result.
The metric (\ref{ss}) produces always a larger effect than the metric (\ref{sls}) reflecting that in the Gauss's mode a probe-mass, treated as a point-like object, loses part of the interaction effects because it cannot reach all the field  isotropically smeared around the central mass.\\
Just as a curiosity we mention that we can alternatively see (\ref{sls}) as a linearized solution and in this case nothing can be said about $\beta$; it is just missing in (\ref{sls}). But we can use the following artifice: we substitute $r$ by the ``physical" distance $r_{1}$, with 
\be
r_{1}\equiv \sqrt{g_{11}}r=\sqrt{1+\frac{\chi}{3r}}\;r.
\ee
Then, $dr_1\approx dr$ and
\be
\frac{\chi}{r}=\frac{\chi}{r_1}\sqrt{g_{11}}\approx\frac{\chi}{r_1}(1+\frac{\chi}{6r})\approx\frac{\chi}{r_1}+\frac{\chi^2}{6r^{2}_1}+\cdots
\ee
So, the metric (\ref{sls}) becomes
\be
ds^2\approx-(1-\frac{\chi}{r_1}-\frac{\chi^2}{6r^{2}_1}+\cdots)dt^2+(1+\frac{\chi}{3r_1}+\frac{\chi^2}{18r^{2}_1}+\cdots)(dx^2+dy^2+dz^2),
\ee
for which $\alpha=1,\;\beta=-\frac{1}{3},\;\gamma=\frac{1}{3}$. Then we have
\be
\Delta\phi=\Delta\phi_{E},
\ee
showing an impressive but artificial improvement of the fitting.\\
For the bending of a starlight by the Sun we have 
\be
\Delta\theta=\frac{1+\gamma}{2}\Delta\theta_{E}=\frac{2}{3}\Delta\theta_{E}.
\ee
confirming that the Gauss's mode cannot be used for fitting these two-interacting-body measurements.
Let us now consider a case where the Gauss's view must necessarily be implemented. It corresponds to a system with many objects interacting with our probe-mass, all so distant that they can be treated as point-like objects but also, as they are separated by such relatively small distances, they must be considered as an extended set of points; in a limit they can even be seen as forming  a continuous set. This necessarily happens in a cosmological scale due to the approximate homogenous and isotropic mass distribution in the universe. So our probe-mass perceives an isotropic space:  there is indeed at least one mass-probe in every direction ${\vec f}$. This is a perfect arrangement for the Gauss's mode but the resultant effective metric is not (\ref{sls}) which corresponds just to a two-body interaction. In order to see that we should first understand the meaning of the factor 1/3 in (\ref{sls}). It is obviously a consequence of the averaging over the space 3-dimensions.  The metric tensor in (\ref{sls}) (or better, its shift from the Minkowski metric) reflects the interaction field between the two interacting masses, i.e. a flux of classical gravitons along a given null direction whose space component is specified by the two interacting masses. 
When dealing through (\ref{sls}) with a probe-mass interaction we are taking a field that exists only along the straight line segment between the mass probe and the central mass, and we are smearing it isotropically on the 3-space around the central mass. That explains the factor 1/3 on the shift from flat space in the metric space-components. The metric time-component is not affected reflecting the fact that the actual field goes along just one space-direction, or in other words, along a null spacetime direction.
Roughly it means that we are equally distributing on all space directions an effect that exists only along an specific one. This explains the 1/3 factor in (\ref{sls}) and why the Gauss's mode does not fit those classical tests.  

 In a cosmological scale, due to the isotropy of the mass distribution in the universe, we do not smear on the whole space a field that exist only along a particular direction, we are taking the actual mean field, as there is one probe-mass in every direction ${\vec f}$. So. So, we must multiply by 3 all the shifts from flat spacetime in the metric (\ref{sls}), including its time component. Thus we would have
\be
\l{msls}
ds^2=-(1-\frac{3\chi}{r})dt^2+(1-\frac{\chi}{r})(dx^{2}+dy^{2}+dz^{2}),
\ee 
instead of (\ref{sls}). Therefore, in a cosmological scale, a mass $M=\frac{\chi}{2G}$ produces a gravitational acceleration 3 times larger than the one predicted by the Schwarzschild metric (\ref{ss}), or the Newtonian theory for that means. This is amazing because it solves in a very natural way the mass-deficit problem of modern cosmology, according to which about 2/3 of the necessary mass required by general relativity to fit the present spacetime curvature is missing from the observed data.\\
Our point is that General Relativity must be seen as a microscopic theory defined on a line $f$ by extended causality \cite{gr-qc/9801040}. Its continuous macroscopic solutions are average-valued fields whose definitions are sensitive to the experimental method employed on their detections.
All interactions are actually described by discrete, ``quantized" fields but in general (except in some cases like in the photoelectric effect and in the Compton scattering) this is not reflected in the output of our measuring apparatus as they just detect spacetime averaged effects. So, what is actually measured are effective fields whose definitions depends on the kind of measurements involved. For the gravitational field this is well exemplified in the figure below. For point-mass-point-mass interactions the Coulomb's view, i.e. the Schwarzschild solution (\ref{ss}), is the most appropriate description as the Gauss's view would produce just a fraction of the actual field, represented by the dotted circle in (a). But for interactions with a continuous and isotropic mass distribution, as it happens in a cosmological scale, the Gauss's mode (b) is the best suited one. 
\vglue-2cm
\begin{figure}
\l{gvc}
\vglue1cm
\epsfxsize=200pt
\epsfbox{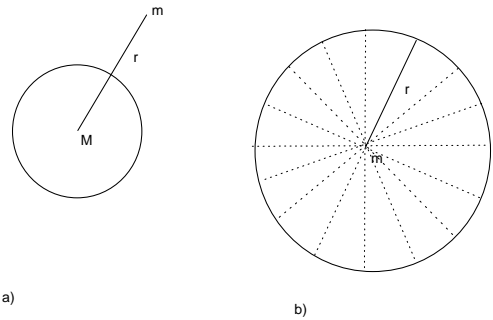}

\vglue-3cm

\caption[Gauss's vs Coulomb's views]{a) A point-mass--point-mass interaction is fitted by the Coulomb's view (the Schwarzschild metric) as the Gauss's view would miss part of the effective field. b) The space isotropy, in a cosmological scale, implies on a continuous mass distribution with each direction corresponding to at least one point-mass. Field measurements in such a context are best fitted by the Gauss's view.}
\parbox[]{15.0cm}{}
\vglue-.5cm
\end{figure}

Despite being defined in a classical context this stricter causality concept of a discrete field formalism does not allow radiative corrections and vacuum fluctuations. The cosmological constant must be zero, the spacetime curvature in a cosmological scale is close to zero and the existing mass in the Universe produces a gravitational acceleration 3 times larger than the one predicted by the Coulomb's (Schwarzschild) view.

\acknowledgments

R. N. Silveira acknowledges a grant from CAPES for writing his M.Sc. dissertation.

\end{document}